\documentclass[12pt,epsf,amssymb]{article}

\usepackage{graphicx}
\usepackage{axodraw}
\usepackage{amsfonts}
\usepackage{amsmath}
\usepackage[usenames]{color}
\usepackage{amssymb}

\def\Z{\mathbb{Z}}

\begin{document}
\baselineskip 0.6cm
\newcommand{\vev}[1]{ \left\langle {#1} \right\rangle }
\newcommand{\bra}[1]{ \langle {#1} | }
\newcommand{\ket}[1]{ | {#1} \rangle }
\newcommand{\Dsl}{\mbox{\ooalign{\hfil/\hfil\crcr$D$}}}
\newcommand{\nequiv}{\mbox{\ooalign{\hfil/\hfil\crcr$\equiv$}}}
\newcommand{\nsupset}{\mbox{\ooalign{\hfil/\hfil\crcr$\supset$}}}
\newcommand{\nni}{\mbox{\ooalign{\hfil/\hfil\crcr$\ni$}}}
\newcommand{\EV}{ {\rm eV} }
\newcommand{\KEV}{ {\rm keV} }
\newcommand{\MEV}{ {\rm MeV} }
\newcommand{\GEV}{ {\rm GeV} }
\newcommand{\TEV}{ {\rm TeV} }

\def\diag{\mathop{\rm diag}\nolimits}
\def\tr{\mathop{\rm tr}}

\def\Spin{\mathop{\rm Spin}}
\def\SO{\mathop{\rm SO}} 
\def\O{\mathop{\rm O}}
\def\SU{\mathop{\rm SU}}
\def\U{\mathop{\rm U}}
\def\Sp{\mathop{\rm Sp}}
\def\SL{\mathop{\rm SL}}

\def\change#1#2{{\color{blue}#1}{\color{red} #2}\color{black}\hbox{}}
%\def\change#1#2{#2}

%%%%%%%%%%
%%%%%%%%%%      title page
%%%%%%%%%%

\begin{titlepage}

\begin{flushright}
LTH 705 \\
UCB-PTH-06/08 \\
LBNL-60200 \\
\end{flushright}

\vskip 2cm
\begin{center}
{\large \bf A Stable Proton without $R$ Parity \\
  --- Implications for the LSP ---} 

\vskip 1.2cm
${}^a$Radu Tatar and ${}^b$Taizan Watari

\vskip 0.4cm

${}^a${\it Division of Theoretical Physics, Department of Mathematical Sciences

The University of Liverpool,
Liverpool,L69 3BX, England, U.K.

rtatar@liverpool.ac.uk}

${}^b${\it Department of Physics and Lawrence Berkeley National 
Laboratory,

University of California, Berkeley, CA 94720, USA

TWatari@lbl.gov} \\

\vskip 1.5cm

\abstract{The proton decays too rapidly in supersymmetric theories if a 
dimension-4 operator $\bar{\bf 5} \cdot {\bf 10} \cdot \bar{\bf 5}$ exists 
in the superpotential. The conventional idea is to impose the $R$-parity 
to kill this operator with a stable lightest supersymmetry 
particle (LSP) as a direct consequence. However, the SUSY-zero mechanism is also able to kill 
the operator without an unbroken $R$-parity. 
In this article, we provide a firm theoretical justification for the 
absence of the dimension-4 proton decay operator under the SUSY-zero 
mechanism, by using some input from string theory. 
The LSP may be unstable without the $R$-parity and, indeed, some dimension-5 
$R$-parity violating operators may be generated in effective theories. This 
suggests that the dark matter is an axion in this string theory inspired 
model. An insight on the SUSY-zero mechanism is also obtained.
}

\end{center}
\end{titlepage}

%%%%%%%%%%
%%%%%%%%%%      main part 
%%%%%%%%%%

%%%%%%%%%%%%%%%%%%%%%%%%%%%%%%%%%%%%%%%%%%%%%%%%%%%%%%%%%%
%\section{Introduction}
%%%%%%%%%%%%%%%%%%%%%%%%%%%%%%%%%%%%%%%%%%%%%%%%%%%%%%%%%%

The SU(5)$_{\rm GUT}$ gauge coupling unification of supersymmetric extensions of 
the standard model is quite remarkable. Supersymmetric theories, however, 
allow dimension-4 operators that break baryon number and lepton number, 
and lead to too rapid proton decay. 

The dimension-4 proton decay operators 
\begin{equation}
W \ni \bar{D} \cdot \bar{U} \cdot \bar{D} + \bar{D} \cdot Q\cdot L 
  + L \cdot \bar{E} \cdot L
\label{eq:dim4-MSSM}
\end{equation}
are simply written as 
\begin{equation}
  W \ni \bar{\bf 5} \cdot {\bf 10} \cdot \bar{\bf 5}
\label{eq:dim4}
\end{equation}
in terms of Georgi--Glashow SU(5)$_{\rm GUT}$ unified multiplets. 
Conventional idea has been to impose a matter parity or $R$ parity to kill these
operators. Chiral multiplets ${\bf 10}$ and $\bar{\bf 5}$ are odd and 
$H({\bf 5})$ and $\bar{H}({\bar{\bf 5}})$ are even under the matter parity.
The dimension-4 proton decay operator (\ref{eq:dim4}) is odd under 
the parity and vanishes in a theory with the matter parity. 
Since the $R$ parity is just a combination of the matter parity and $(-1)^F$, 
where $F$ is the Fermion number, the $R$ parity is equivalent to the matter 
parity. The lightest supersymmetry particle (LSP) is stable 
in a theory with an unbroken $\Z_2$ $R$ (and matter) parity \cite{review}.

The dimension-4 operator (\ref{eq:dim4}) is absent due to the $\Z_2$ matter 
parity but this argument cannot be turned around and the absence of the operator 
(\ref{eq:dim4}) is not enough to conclude that 
there is a $\Z_2$ symmetry and that the LSP is stable. Indeed, 
the operator (\ref{eq:dim4}) is absent in the framework proposed in 
\cite{TW}, although the $\Z_2$ matter parity is broken.
In this article, we rederive the absence of (\ref{eq:dim4}) 
with a plain D = 4 field theory language.
We further study the phenomenological consequences of the absence of 
the $R$ parity, with the LSP decay a particular case.
An insight on the SUSY-zero mechanism is also obtained as a bi-product. 

The dimension-4 proton decay operator (\ref{eq:dim4}) is absent in \cite{TW}
essentially because of the SUSY-zero mechanism.
Let us consider two SU(5)$_{\rm GUT}$~$\times$~U(1) gauge theories 
with U(1) charges of the chiral multiplets given in Table \ref{tab:charge}. 
Those D = 4 field-theory models are simplified versions of models in 
\cite{TW}; ingredients essential to the absence of (\ref{eq:dim4}) are 
not lost in the simplification.\footnote{The 
%WE discuss how the D = 4 field-theory model captures the essence \cite{TW}.
SU(5)$_{\rm GUT}$~$\times$~U(1) gauge group is embedded in $G=E_7$ or $E_8$ 
in string theory in \cite{TW}. See the appendix for more about the relation 
between the models in \cite{TW} and those presented in this article.}
Chiral multiplets $\bar{N}^c$ are in the Hermitian conjugate representation 
of the right-handed neutrino chiral multiplets $\bar{N}$. The U(1) gauge 
symmetry is not necessarily free of anomaly; the anomaly is cancelled 
by the generalized Green--Schwarz mechanism. 
The $\Z_5$ subgroup of the U(1) gauge symmetry of both models is equivalent 
to the $\Z_5$ centre of the SU(5)$_{\rm GUT}$ symmetry, and $\Z_{10}$ subgroup 
of the U(1) symmetries gives the matter parity $\Z_{10}/\Z_5 \simeq \Z_2$. 
%%%%%%%%%%%%%%%%%%%%%%%%%%%%%%%%%%%%%%%%%%%%%%%%%%%%%%%%%%%%%%%%%%%
\begin{table}[tb]
\begin{center}
\begin{tabular}{|r|c|c|c|c|c|c|c|}
\hline
Chiral multiplets & & {\bf 10} & $\bar{\bf 5}$ & $H({\bf 5})$ &
   $\bar{H}(\bar{\bf 5})$ & $\bar{N}$ & $\bar{N}^c$ \\
\hline
U(1) Charge & 4+1 model & $-1$ & $3$ & $2$ & $-2$ & $-5$ & $5$ \\
\hline
U(1) Charge & 3+2 model & $-3$ & $-1$ & $6$ & $4$ & $-5$ & $5$ \\
\hline
\end{tabular}
\caption{\label{tab:charge} U(1)-charge assignments of (the simplified version 
of) the two models of \cite{TW}. These charge assignments allow all of 
Dirac neutrino Yukawa couplings 
$W \ni \bar{N} \cdot \bar{\bf 5} \cdot H({\bf 5})$, up-type, down-type and 
charged-lepton Yukawa couplings. Chiral multiplets with $^{c}$ are in the 
Hermitian conjugate representation of those without $^{c}$
under the SU(5)$_{\rm GUT}$~$\times$~U(1).}
\end{center}
\end{table}
%%%%%%%%%%%%%%%%%%%%%%%%%%%%%%%%%%%%%%%%%%%%%%%%%%%%%%%%%%%%%%%%%%%%
In the 4+1 model, the dimension-4 proton decay operator (\ref{eq:dim4}) is 
absent even if chiral multiplets $\bar{N}^c$ have non-zero expectation values, 
because any operators of the form 
\begin{equation}
 W \nni \bar{\bf 5} \cdot {\bf 10} \cdot \bar{\bf 5} \cdot
  \vev{\bar{N}^c}^{n \geq 0}
\label{eq:41-vanish}
\end{equation}
are forbidden by the U(1) gauge symmetry. Likewise in the 3+2 model, 
some of chiral multiplets $\bar{N}$ may have non-zero expectation values, 
yet (\ref{eq:dim4}) is absent, because 
\begin{equation}
 W \nni \bar{\bf 5} \cdot {\bf 10} \cdot \bar{\bf 5} \cdot \vev{\bar{N}}^{n
  \geq 0}
\label{eq:32-vanish}
\end{equation}
are not allowed by the U(1) gauge symmetry. This is so-called the 
SUSY-zero mechanism; when a U(1) symmetry is broken by vacuum expectation 
values (vev's) of positively [negatively] charged chiral multiplets, 
operators in the superpotential that appear to have negative [positive, 
respectively] charge are allowed because the vev's may supply the appropriate 
U(1) charge; but operators that appear to have positive [negative, resp.] 
U(1) charge---like (\ref{eq:41-vanish}) [(\ref{eq:32-vanish}), resp.] are not, 
because the multiplets with the vev's cannot supply negative [positive, resp.] 
U(1) charge so that the operators become U(1)-invariant.
The vev's of the chiral multiplets $\bar{N}^c$ or $\bar{N}$ break the matter 
parity, yet the dimension-4 proton decay operators (\ref{eq:dim4}) are absent. 
Thus, the SUSY-zero mechanism can be an alternative to the matter parity.
%condition that a continuous U(1) symmetry is unbroken plays the role of 
%matter parity in eliminating certain terms in the superpotential.

The above argument, however, crucially depends on an assumption that the vev's 
are inserted only in non-negative power as in (\ref{eq:41-vanish}) and 
(\ref{eq:32-vanish}). 
In supersymmetric quantum field theories, in general, superpotential can be
arbitrary, as long as it is holomorphic in chiral multiplets. It can also 
have a pole or singularity, on a K\"{a}hler manifold parametrized by chiral 
multiplets. Thus, it is hard to justify the absence of operators with $n < 0$ 
only with ${\cal N} = 1$ supersymmetry of D = 4 field theories. 
In effective field theories that arise from geometric compactification of 
string theory, however, we see in the following that we have a better answer 
to this question: the SUSY-zero mechanism is justified for renormalizable 
operators, though not necessarily for non-renormalizable operators. 
In particular, restricting to $n \geq 0$ in renormalizable operators 
(\ref{eq:41-vanish}) and (\ref{eq:32-vanish}) will be justified.

Heterotic $E_8 \times E_8$ string theory has a superpotential \cite{super}
\begin{equation}
W \ni \int d^6 z \; \Omega \wedge \tr {}_{E_8 \mbox{-}{\rm adj.}}
\left(A dA - i \frac{2}{3}AAA \right).
\label{eq:Het-super}
\end{equation}
This describes a part of super Yang--Mills interactions. 
There are 16 supersymmetry charges locally, and these large supersymmetry 
and gauge symmetry constrain the superpotential to the form (\ref{eq:Het-super}); 
note that it stops at the cubic term, and the $\alpha'$ corrections, vanish 
\cite{Witten}.\footnote{Although non-perturbative effects of string theory 
such as world-sheet instantons can generate extra contributions 
to the superpotential, we ignore them because their effects can be small.}  
After compactification, this superpotential is re-written in terms of 
infinite number of D = 4 chiral multiplets. 
The superpotential (\ref{eq:Het-super}) decomposes into supersymmetric 
mass terms and tri-linear interactions of chiral multiplets. 
Depending on which part of $\mathfrak{e}_8/\mathfrak{su}(5)_{\rm GUT}$ 
each chiral multiplet comes from, its representation under 
the SU(5)$_{\rm GUT} \times$U(1) gauge group is different 
(see Table \ref{tab:charge} and the appendix). 
We call them U(1) eigenstates. In terms of the U(1) eigenstates, 
the superpotential (\ref{eq:Het-super}) consists of tri-linear interactions 
\begin{eqnarray}
 W & \ni & (y_u)_{ijk} {\bf 10}_i . {\bf 10}_j . H({\bf 5})_k + 
    (y_{d,e})_{ijk} \bar{\bf 5}_i \cdot {\bf 10}_j \cdot \bar{H}(\bar{\bf 5})_k 
  + (y_{\nu})_{ijk} \bar{N}_i \cdot \bar{\bf 5}_j \cdot H({\bf 5})_k \\
   & + & (y_u^c)_{ijk} {\bf 10}^c_i . {\bf 10}^c_j . H({\bf 5})^c_k 
     +   (y_{d,e}^c)_{ijk} \bar{\bf 5}^c_i \cdot {\bf 10}^c_j \cdot 
                           \bar{H}(\bar{\bf 5})^c_k 
     + (y_{\nu}^c)_{ijk} \bar{N}^c_i \cdot \bar{\bf 5}^c_j \cdot H({\bf 5})^c_k 
   \nonumber
\label{eq:Yukawa}
\end{eqnarray}
and supersymmetric mass terms
\begin{eqnarray}
 W & \ni & (M_{\bar{\bf 5}})_{ij} \bar{\bf 5}_i \cdot \bar{\bf 5}^c_j 
   + (M_{\bar{H}})_{ij} \bar{H}(\bar{\bf 5})_i \cdot \bar{H}(\bar{\bf 5})^c_j \\
  & &   + (M_{{\bf 10}})_{ij} {\bf 10}_i \cdot {\bf 10}^c_j 
   + (M_{\bar{N}})_{ij} \bar{N}_i \cdot \bar{N}^c_j
   \; \left[ + (M_H)_{ij} H({\bf 5})_i \cdot H({\bf 5})^c_j \right] + \cdots.
  \nonumber
\label{eq:SUSY-mass}
\end{eqnarray}
Indices $i,j,k$ label infinite particles in the Kaluza--Klein tower of 
U(1) eigenstates.
%Other supersymmetric mass terms are omitted because only the mixing 
%between $\bar{\bf 5}$-type and $\bar{H}(\bar{\bf 5})$-type U(1) eigenstates
%is relevant in the absence or presence of the dimension-4 proton decay 
%operator. 
Note that $\bar{H}(\bar{\bf 5})^c$-type U(1) eigenstates are nothing but 
the $H({\bf 5})$-type eigen states in the 4+1 model, 
but those two classes of states are different in the 3+2 model (see 
Table~\ref{tab:charge} and the appendix).
The number of {\bf 10}-type chiral multiplets should be 
larger than that of ${\bf 10}^c$-type by 3, corresponding to the 
three generations of $(\bar{U},Q,\bar{E})$. Similar chirality constraint 
exists for the multiplets in the SU(5)$_{\rm GUT}$-{\bf anti.-fund.} and 
{\bf fund.} representations; 
% in the 4+1 model for example, the total number of chiral multiplets 
% of $\bar{\bf 5}$-type and $\bar{H}(\bar{\bf 5})$-type is larger than 
% that of $\bar{\bf 5}^c$-type and $H({\bf 5})$-type by 3, so that 
% we have 3 generations of down-type quarks and lepton doublets. 
String theory provides a dictionary translating topological information 
into this net chirality. Here, we just assume that the compactification 
geometry is chosen, so that the net chirality of the real world is reproduced.
There is an additional constraint on the rank of the mass matrices, so that   
we have the electroweak Higgs doublets in the low-energy spectrum.
The mass eigenvalues of the mass matrices $M_{\bar{\bf 5}}$, $M_{\bar{H}}$ etc. 
are not necessarily either zero or of the order of the Kaluza--Klein scale; 
some mass eigenvalues are determined by moduli parameters of compactification.
There is no way specifying those eigenvalues without further specifying details 
of compactification, and we just leave them as arbitrary parameters of effective 
field theories. The terms of the superpotential (\ref{eq:Yukawa}) and (\ref{eq:SUSY-mass}),
preserves the U(1) symmetry, as it should be.

As long as neither $\bar{N}$ nor $\bar{N}^c$ has non-zero expectation 
values, the U(1) symmetry is not broken. The distinction between 
the $\bar{\bf 5}$-type and $\bar{H}(\bar{\bf 5})$-type U(1) eigenstates
is maintained, and the dimension-4 proton decay operator (\ref{eq:dim4}) is 
forbidden by the U(1) symmetry. This is the place we start off, and we examine 
how the expectation values of $\bar{N}$ or $\bar{N}^c$ would affect the 
low-energy effective field theories. 

The superpotential of low-energy effective theory is obtained by i) identifying 
the massless modes, and ii) integrating out all but massless modes from the 
theory. Instead of dealing with infinite D = 4 chiral multiplets to be integrated
out, we consider simpler models, where only finitely many Kaluza--Klein 
particles are maintained, so that we can deal with finite-by-finite mass 
matrices, instead of infinite-by-infinite ones. This is the approximation we are going to consider in the 
present paper. 

When either $\bar{N}^c$ [or $\bar{N}$] develops a non-zero expectation value, 
the sixth term [or the third term] in (\ref{eq:Yukawa}) gives rise to a 
deformation in the supersymmetric mass matrix. Since a part of the mass matrices
carry a non-zero U(1)-charge, mass eigenstates are no longer pure U(1) 
eigenstates. In particular, massless states 
in the SU(5)$_{\rm GUT}$-{\bf anti.fund.} representation are no longer expected 
to be either pure $\bar{\bf 5}$-type or pure $\bar{H}(\bar{\bf 5})$-type 
U(1) eigenstates. Thus, this is potentially dangerous. 

Let us first examine the massless modes in the simplified version of the 
4+1 model, where only finitely many Kaluza--Klein particles are taken 
into account. Let us consider a model with 
\begin{itemize}
\item 4 $\bar{\bf 5}$-type chiral multiplets, denoted by 
                     $\bar{\bf 5}_i$ ($i=1,\cdots,4$), 
\item 1 $\bar{\bf 5}^c$-type chiral multiplet, 
\item 2 $\bar{H}(\bar{\bf 5})$-type chiral multiplets denoted by 
   $\bar{H}_k$ ($k=1,2$) and 
\item 2 $H({\bf 5})$-type chiral multiplets denoted by $H_l$ ($l=1,2$) 
\end{itemize}
(in addition to 3 {\bf 10}-type chiral multiplets).
The supersymmetric mass matrix of multiplets in the 
SU(5)$_{\rm GUT}$-{\bf fund.} and -{\bf anti.fund.} representations is given by 
\begin{equation}
 W \ni \left( H_l, \bar{\bf 5}^c \right) \left(
	\begin{array}{cc}
	 (M_H)_{lk} & 0 \\ 
         (y^c_{\nu} \vev{\bar{N}^c} )_{k} & (M_{\bar{\bf 5}})_{i} 
	\end{array}\right)
  \left(\begin{array}{c}
   \bar{H}_k \\ \bar{\bf 5}_i
	\end{array}\right).
\label{eq:41-def-mmat}
\end{equation}
The 2 by 2 matrix $(M_H)_{lk}$ is assumed to be of rank 1, so that 
we have a pair of light Higgs doublets, $H_u$ and $H_d$. 
Here, we have already chosen $\vev{\bar{N}^c} \neq 0$, and set 
$\vev{\bar{N}} = 0$, following \cite{TW}.
Now, without a loss of generality, we can change the basis within 
$H_l$, $\bar{H}_k$ and $\bar{\bf 5}_i$, so that $(M_H)_{lk}=0$ except 
$(M_H)_{22}=M_H \neq 0$, and so that $(M_{\bar{\bf 5}})_{i=1,2,3} = 0$ and 
$(M_{\bar{\bf 5}})_4 = M_{\bar{\bf 5}} \neq 0$. The 3 by 6 mass matrix 
(\ref{eq:41-def-mmat}) becomes 
\begin{equation}
 W \ni \left(H_1,H_2,\bar{\bf 5}^c\right)
  \left(\begin{array}{cccc}
   0 & 0 & 0 & 0 \\ 0 & M_H & 0 & 0 \\ (y^c_\nu
    \vev{\bar{N}^c})_1 & (y^c_\nu \vev{\bar{N}^c})_2 &
    0 & M_{\bar{\bf 5}}
	\end{array}\right)
\left(\begin{array}{c}
 \bar{H}_1 \\ \bar{H}_2 \\ \bar{\bf 5}_{1,2,3} \\ \bar{\bf 5}_{4}
      \end{array}\right).
\end{equation}
We are primarily interested in the massless modes, which describe the 
low-energy effective theories. 
Although all the massive states are mixture of U(1) eigenstates, 
the U(1) eigenstates $H_1$ and $\bar{\bf 5}_{1,2,3}$ are massless 
(and hence, mass eigen-) states, as well.
Those low-energy multiplets, denoted by $\hat{}$ on it, namely 
$\hat{\bar{{\bf 5}}}_{1,2,3} = \bar{\bf 5}_{1,2,3}$ and 
$\hat{H} = H_1$ are identified with 3 generations of $(\bar{D},L)$ 
and a quintet containing $H_u$, respectively.
The other massless chiral multiplet, $\hat{\bar{{\bf 5}}}_0$, is given by 
a linear combination $\propto \left(M_{\bar{\bf 5}} \bar{H}_1 
- (y^c_\nu \vev{\bar{N}^c})_1 \bar{\bf 5}_4 \right)$. This is 
to be identified with a quintet containing $H_d$.
Thus, an important observation is that some of massless modes still remain 
to be U(1) eigenstates, although all the massive states are mixture of 
U(1) eigenstates. 

It is important, in particular, that all three massless states 
$\hat{\bar{{\bf 5}}}_{1,2,3}=(\bar{D},L)_{1,2,3}$ remain to be pure 
$\bar{\bf 5}$-type U(1) eigenstates. When the superpotential (\ref{eq:Yukawa}) 
written in terms of U(1) eigenstates are re-written in terms of mass eigenstates,
and when terms that only involve massless states are retained, 
the U(1) eigenstates $\bar{\bf 5}_i$ turn into both 
$\hat{\bar{{\bf 5}}}_{1,2,3} = (\bar{D},L)_{1,2,3}$ and 
$\hat{\bar{{\bf 5}}}_0 = H_d$, but $\bar{H}_k$ turn only into 
$\hat{\bar{{\bf 5}}}_0 = H_d$, not to $\hat{\bar{{\bf 5}}}_{1,2,3}$. 
% By reversing the relation, the U(1) eigenstates $\bar{H}_1$ and 
% $\bar{\bf 5}_4$ contains the $\hat{\bar{{\bf 5}}}_0=H_d$ component.
% When the superpotential (\ref{eq:Het-super}) written in terms of 
% U(1) eigenstates, $\bar{\bf 5}_i$, $\bar{\bf 5}^c$, 
% $\bar{H}(\bar{\bf 5})_k$ and $H({\bf 5})_l$ is rewritten in terms of 
% mass eigen states, and when terms involving only the massless states 
% are retained, 
Thus, the dimension-4 operator (\ref{eq:dim4}) that involves two of 
$\hat{\bar{{\bf 5}}}_{1,2,3} = (\bar{D},L)_{1,2,3}$ does not arise 
from the super Yang--Mills interaction (\ref{eq:Yukawa}) and 
(\ref{eq:SUSY-mass}), even when $\vev{\bar{N}^c} \neq 0$ 
and the matter parity is broken. 
This result has been obtained in \cite{TW}, although phrased in a more 
geometric language (see also the appendix). Note also that a lepton number 
violating operator 
\begin{equation}
W \ni y_{d,e} \vev{\bar{N}^c} H_d \cdot \bar{E} \cdot H_d
\end{equation}
could have been generated, but it vanishes when there is only on doublet 
because of anti-symmetric contraction of SU(2) indices.

Let us continue the analysis a little further, before drawing a conclusion 
on the dimension-4 proton decay operator. The same analysis can be carried 
out,\footnote{There, one has to assume that the 2 by 6 matrix 
$((M_H)_{lk},(y_\nu \vev{\bar{N}})_{li})$ is of rank 1.}  
now assuming $\vev{\bar{N}} \neq 0$ and $\vev{\bar{N}^c} = 0$, 
instead of the other way around. This case should lead to dimension-4 
proton decay \cite{TW}, and let us confirm it in the D = 4 field-theory 
language we used above. We can see by diagonalizing the mass matrix that 
massive states are mixture of U(1) eigenstates, but some of massless states 
remain to be pure U(1) eigenstates, just as in the previous case with 
$\vev{\bar{N}^c} \neq 0, \vev{\bar{N}} = 0$. The massless states 
$\hat{H} = H_1 \supset H_u$ and $\hat{\bar{{\bf 5}}}_{1,2}=(\bar{D},L)_{1,2}$ 
still remain pure U(1) eigenstates, just as in the previous case.  
The difference from the previous case is that $\hat{\bar{H}} = \bar{H}_1 
\supset H_d$ becomes a pure $\bar{H}(\bar{\bf 5})$-type U(1) eigenstate, and 
the other massless state 
$\hat{\bar{{\bf 5}}}_{3} = (\bar{D},L)_3 \propto \left(M_H \bar{\bf 5}_3 
- (y_\nu \vev{\bar{N}})_3 \bar{H}_2 \right)$ becomes the mixture of U(1) 
eigenstates, instead. When converting the superpotential 
(\ref{eq:Yukawa})+(\ref{eq:SUSY-mass}) in the U(1) eigenbasis into 
the mass eigenbasis and retaining only the massless states, $\bar{H}_2$ 
contains the $\hat{\bar{{\bf 5}}}_3$ component, and the second term of 
(\ref{eq:Yukawa}) becomes 
\begin{equation} 
  y_{d,e} \, \bar{\bf 5}_{1,2,3} \cdot {\bf 10} \cdot \bar{H}_{1,2}
  \rightarrow 
   y_{d,e} \, \hat{\bar{{\bf 5}}}_{1,2} \cdot {\bf 10} \cdot H_d + 
   y_{d,e} \,  \hat{\bar{{\bf 5}}}_3 
\cdot {\bf 10}  \cdot H_d -
   y_{d,e} \, \hat{\bar{{\bf 5}}}_{1,2} \cdot {\bf 10} \cdot 
      \left(\frac{y_\nu \vev{\bar{N}}}
                 {\sqrt{|M_H|^2+|(y_\nu \vev{\bar{N}})|^2}} 
      \right) \hat{\bar{{\bf 5}}}_3.
\end{equation}
Thus, the dimension-4 proton decay operator is indeed generated when 
$\vev{\bar{N}} \neq 0$. The $\vev{\bar{N}}$-dependence of the last 
term explains why the single $\vev{\bar{N}}$ insertion captures the 
physics at the level of whether certain operators vanish or not, even 
when $\vev{\bar{N}} \gg M_H$. The single $\vev{\bar{N}}$ insertion was 
also used in the discussion in \cite{TW}, where it was backed by 
a geometric intuition. Now we have an independent confirmation of how and 
why it works. 

So far, we have discussed only renormalizable (dimension-4) operators 
in the low-energy effective theories. The renormalizable part of the 
low-energy effective superpotential is given by re-writing the 
U(1) eigenstates in terms of mass eigenstates and just drop all the 
terms containing heavy states \cite{Witten}. We have seen in both cases, 
namely $\vev{\bar{N}^c} \neq 0$ and $\vev{\bar{N}} \neq 0$,  
in the D = 4 field theory language (and in a more geometric language 
in \cite{TW}) that the mixing between the U(1) eigenstates can be 
traced in the low-energy effective superpotential by single ($n=1$) 
insertion of the expectation values $\vev{\bar{N}^c}$ and 
$\vev{\bar{N}}$. As a consequence, we have seen that $\vev{\bar{N}} \neq 0$ 
does generate dimension-4 proton decay operator (\ref{eq:dim4}), but 
the low-energy effective theories remain free of (\ref{eq:dim4}) as long as 
$\vev{\bar{N}} = 0$, even if $\vev{\bar{N}^c} \neq 0$ and the matter parity 
is not preserved. Similar analyses can be carried out for the 3+2 model, 
but it is essentially the same as in the 4+1 model, and we do not repeat here.

We have so far implicitly assumed that the 
$\bar{\bf 5}$--$\bar{H}(\bar{\bf 5})$ mixing arises only from 
the superpotential. This is the case if the K\"{a}hler potential is of 
the form 
\begin{eqnarray}
K & = & Z^{\bar{\bf 5}}_{ij} \bar{\bf 5}^\dagger_i \bar{\bf 5}_j
      + Z^{\bar{H}}_{ij} \bar{H}^\dagger_j \bar{H}_j  \\
& + & c_{ijkl} \bar{\bf 5}_i^\dagger \bar{\bf 5}_j^\dagger
               \bar{\bf 5}_k \bar{\bf 5}_l
 + c'_{ijkl} \bar{H}_i^\dagger \bar{H}^\dagger_j \bar{H}_k \bar{H}_l 
 + c''_{ijkl} \bar{\bf 5}_i^\dagger \bar{\bf 5}_j \bar{N}^\dagger_k \bar{N}_l
 + \cdots.
\label{eq:safe-Kahler}
\end{eqnarray}
However, U(5)$_{\rm GUT}$~$\times$~U(1) invariant K\"{a}hler potential 
of ${\cal N} = 1$ supersymmetry may have such terms as 
\begin{equation}
K \ni \kappa_n \bar{\bf 5}^\dagger \bar{N}^c \bar{H} 
         (\bar{N}^{c \dagger} \bar{N}^c)^n + 
      \lambda_n \bar{\bf 5}^\dagger \bar{N}^\dagger \bar{H} 
         (\bar{N}^\dagger \bar{N})^n  + {\rm h.c.},
\label{eq:mix-Kahler}
\end{equation}
which leads to kinetic $\bar{\bf 5}$--$\bar{H}$ mixing when either 
$\vev{\bar{N}^c}$ or $\vev{\bar{N}}$ is non-zero. 
%It is unclear whether the K\"{a}hler potential of effective theories 
%really include contributions as (\ref{eq:mix-Kahler}).
%
%\begin{equation}
% {\cal L} \ni \alpha' \tr \left(F_{LM}F^{MN}F_{N}^{\; L} \right)
%\label{eq:alpha-prime}
%\end{equation}
%
%give rise to some terms in (\ref{eq:mix-Kahler}), but it is known that 
%the $\alpha'$ correction (\ref{eq:alpha-prime}) does not exist in Heterotic 
%%string theory \cite{BdR}. Even if so, such terms as (\ref{eq:alpha-prime}) %
%may exist in other corner of string moduli space, or (\ref{eq:mix-Kahler}) 
%may be generated in the effective theory in the process of integrating out 
%heavy states. Thus, we still have a concern if the arguments so far 
%may not be valid if the K\"{a}hler potential has such terms as 
%(\ref{eq:mix-Kahler}).
Even if the K\"{a}hler potential contains those terms, it turns out that they 
are not a problem. We can re-define the chiral multiplets as 
\begin{equation}
 \bar{\bf 5}' = \bar{\bf 5} + (Z^{\bar{\bf 5}})^{-1} \left( 
    \kappa_n \vev{\bar{N}^c} |\vev{\bar{N}^c}|^{2n} \bar{H}
  + \lambda_n \vev{\bar{N}}^* |\vev{\bar{N}}|^{2n} \bar{H} \right),
\end{equation}
so that the bi-linear part of the K\"{a}hler potential is like 
(\ref{eq:safe-Kahler}) in the newly defined chiral multiplets. 
The superpotential should also be re-written at the same time; 
$\bar{\bf 5}$ in the superpotential is replaced by $\bar{\bf 5}' 
- Z^{-1}(\kappa_0 \vev{\bar{N}^c} + \lambda_0 \vev{\bar{N}} + \cdots)\bar{H}$.
Thus, operators of the form $W \ni \bar{H} \cdot {\bf 10} \cdot \bar{H}$ are 
generated, but as long as we have only one chiral multiplet that have 
the same properties as that of $H_d$ in the low-energy spectrum, 
this operator vanishes because it is anti-symmetric under the exchange of two
$\bar{H}$'s. Thus, the whole arguments for the absence of dimension-4 proton 
decay operator (\ref{eq:dim4}) are not affected, even when the K\"{a}hler 
potential has terms like (\ref{eq:mix-Kahler}).

Now that we have seen that the framework proposed in \cite{TW} guarantees 
that the dangerous dimension-4 operator (\ref{eq:dim4}) is absent, let us 
move on to discuss non-renormalizable operators. Renormalizable (dimension-4) 
terms of the effective superpotential is obtained by truncating all the terms 
that involve heavy states, but the heavy states should not be just truncated, 
but should be integrated out to obtain the superpotential of the effective 
theories. Lots of non-renormalizable operators are generated, in general, 
when heavy states are integrated out.\footnote{Renormalizable terms are not 
affected by this process \cite{Witten}.}

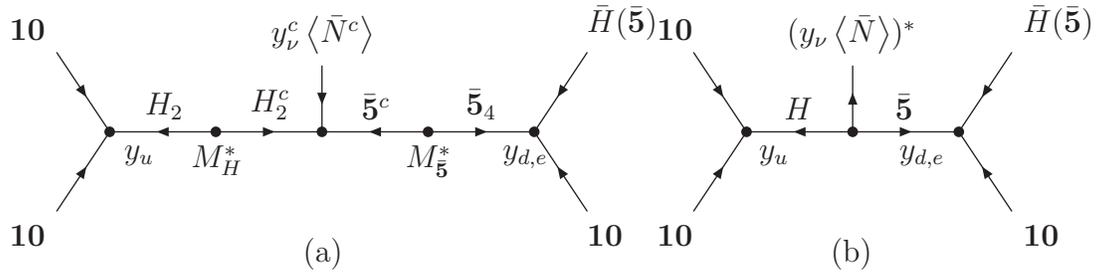
\begin{figure}
\begin{center}
\begin{picture}(400,100)(-200,0)
\ArrowLine(-180,80)(-160,50) \Text(-185,85)[rb]{{\bf 10}}
\ArrowLine(-180,20)(-160,50) \Text(-185,15)[rt]{{\bf 10}}
\Vertex(-160,50){2}
\Text(-155,45)[lt]{$y_u$}
\ArrowLine(-120,50)(-160,50) \Text(-140,55)[b]{$H_2$}
\Vertex(-120,50){2}
\Text(-120,45)[t]{$M_H^*$}
\ArrowLine(-120,50)(-80,50)  \Text(-100,55)[b]{$H^c_2$}
\Vertex(-80,50){2}
\ArrowLine(-80,75)(-80,50)
\Text(-80,80)[b]{$y^c_\nu \vev{\bar{N}^c}$}
\ArrowLine(-40,50)(-80,50)   \Text(-60,55)[b]{$\bar{\bf 5}^c$}
\Vertex(-40,50){2}
\Text(-40,45)[t]{$M_{\bar{\bf 5}}^*$}
\ArrowLine(-40,50)(0,50)  \Text(-20,55)[b]{$\bar{\bf 5}_4$}
\Vertex(0,50){2}
\Text(5,45)[rt]{$y_{d,e}$}
\ArrowLine(20,80)(0,50) \Text(20,85)[bl]{$\bar{H}(\bar{\bf 5})$}
\ArrowLine(20,20)(0,50) \Text(20,15)[tl]{{\bf 10}}
\Text(-80,10)[t]{(a)}
\ArrowLine(60,80)(80,50) \Text(60,85)[br]{{\bf 10}}
\ArrowLine(60,20)(80,50) \Text(60,15)[tr]{{\bf 10}}
\Vertex(80,50){2}
\Text(85,45)[lt]{$y_u$}
\ArrowLine(120,50)(80,50) \Text(100,55)[b]{$H$}
\Vertex(120,50){2}
\ArrowLine(120,50)(120,75)
\Text(120,80)[b]{$(y_\nu \vev{\bar{N}})^*$}
\ArrowLine(120,50)(160,50) \Text(140,55)[b]{$\bar{\bf 5}$}
\Vertex(160,50){2}
\Text(155,45)[tr]{$y_{d,e}$}
\ArrowLine(180,80)(160,50) \Text(185,85)[bl]{$\bar{H}(\bar{\bf 5})$}
\ArrowLine(180,20)(160,50) \Text(185,15)[tl]{{\bf 10}}
\Text(120,10)[t]{(b)}
\end{picture}
\caption{\label{fig:dim5}Super Feynman diagrams for the LSP decay 
operators.}
\end{center}
\end{figure}

Instead of trying to be general, let us look at explicit examples.
Figure \ref{fig:dim5} (a) is a super Feynman diagram, showing that 
\begin{equation}
 W \ni \frac{y_u y_{d,e} y^c_{\nu} \vev{\bar{N}^c}}{M_{\bar{\bf 5}} M_H} \; 
   {\bf 10} \cdot {\bf 10} \cdot {\bf 10} \cdot \bar{H}(\bar{\bf 5})
\end{equation}
is generated in the 4+1 model, if there are massive vector-like pairs 
$(H_2,\bar{H}_2)$ and $(\bar{\bf 5}^c,\bar{\bf 5}_4)$. 
One can also check that this operator is neutral under the U(1)-charge 
assignment of the 4+1 model in Table \ref{tab:charge}. 
$M_H,\; M_{\bar{\bf 5}} \gg \vev{\bar{N}^c}$ is assumed in the coefficient.
Another example is Figure \ref{fig:dim5} (b), where we see that 
\begin{equation}
 W \ni \frac{y_u y_{d,e}}{y_\nu \vev{\bar{N}}} \; 
   {\bf 10} \cdot {\bf 10} \cdot {\bf 10} \cdot \bar{H}(\bar{\bf 5})
\label{eq:in3+2}
\end{equation}
can be generated in the 3+2 model, if there is an extra pair of 
$\bar{\bf 5}$-type and $H({\bf 5})$-type chiral multiplet. 
This operator is also neutral under the U(1)-charge assignment of the 
3+2 model in Table \ref{tab:charge}. It is worth noting that the expectation 
value breaking the anomalous U(1) gauge symmetry can appear not only 
in the numerator but also sometimes in the denominator. 
In the latter example, the vector-like pair of chiral multiplets have 
a mass term only through the Dirac neutrino Yukawa coupling involving 
the expectation value of $\bar{N}$. When the expectation value vanishes, 
the vector-like pair is massless, and the number of massless modes changes.
The locus of $\vev{\bar{N}} = 0$ is a singular locus of the moduli space 
in the topological sector. In such situation, the expectation value 
breaking the U(1) symmetry appear in the denominator of coefficients 
of non-renormalizable operators in the effective theories.
The negatively charged $\vev{\bar{N}}$ in the denominator supplies 
positive U(1) charges in (\ref{eq:in3+2}), neutralizing the negative 
charge of ${\bf 10} \cdot {\bf 10} \cdot {\bf 10} \cdot \bar{H}$. 
Thus, this operator is not eliminated. 
This example clearly shows that the SUSY-zero mechanism does not 
necessarily work for the non-renormalizable operators. 

The dimension-5 operators above effectively look like 
\begin{equation}
W_{\rm eff.} \ni \frac{1}{M_{\rm eff.}} 
 {\bf 10} \cdot {\bf 10} \cdot {\bf 10} \cdot \bar{H}(\bar{\bf 5})
\label{eq:101010Hbar}
\end{equation}
in low-energy physics, whether it comes from the 4+1 model or from 
the 3+2 model. Further non-renormalizable operators are generated 
by integrating out heavy states, and the effective superpotential does not 
necessarily stop at finite-degree polynomial. 
This operator, however, can be a leading contribution to the new physics 
beyond the supersymmetric standard model, and we work on this operator 
in the rest of this article. It is written in terms of chiral multiplets 
of the minimal supersymmetric standard model.as 
\begin{equation}
 W_{\rm eff.} \ni \frac{1}{M_{\rm eff.}} Q \; Q \; Q \; H_d 
    + \frac{1}{M_{\rm eff.}'} Q \; \bar{U} \; \bar{E} \; H_d
\label{eq:LSPdecayOP}
\end{equation}
Baryon number, lepton number symmetries and the matter parity are 
broken by these dimension-5 operators.
The effective coefficients of the two operators are not necessarily 
exactly the same, because of various SU(5)$_{\rm GUT}$ symmetry breaking 
effects, such as the Wilson line.\footnote{Although (\ref{eq:Yukawa}) 
is written in an SU(5)$_{\rm GUT}$-symmetric way, it was just for brevity 
of notation. The spectra and wave functions of mass eigenstates should be 
different for SU(5)$_{\rm GUT}$ partners, due to the 
SU(5)$_{\rm GUT}$-breaking effects, such as the Wilson line, and hence 
the coefficients in (\ref{eq:Yukawa}) are not actually 
SU(5)$_{\rm GUT}$-symmetric. Thus, $M_{\rm eff.}$ and $M_{\rm eff}'$ are 
not expected to be exactly the same, either.}
The effective energy scale $M_{\rm eff.}$ and $M_{\rm eff.}'$
depend on the tri-linear couplings in (\ref{eq:Yukawa}), the 
expectation value $\bar{N}$ or $\bar{N}^c$, and on expectation values of 
SU(5)$_{\rm GUT}$-singlet moduli fields that affect the mass eigenvalues 
such as $M_H$ and $M_{\bar{\bf 5}}$. Although one could come up with some 
naive order-of-magnitude estimate of the first two,\footnote{One could 
use the order of magnitude of the Yukawa couplings we already know for 
the tri-linear Yukawa couplings; for the naive order-of-magnitude estimate 
for $\vev{\bar{N}^c}$ or $\vev{\bar{N}}$, see \cite{TW}.} the stabilization 
of vector bundle moduli is poorly understood in the current string theory, 
and string theory is not able to make a unique prediction of the effective 
energy scale.

Instead, we constrain the range of those energy scales by phenomenological 
limits. The first operator breaks baryon number, and the latter lepton number.
Since proton decay process has to involve both baryon number violation 
and lepton number violation, proton decay amplitudes from 
(\ref{eq:101010Hbar}) should involve both operators of (\ref{eq:LSPdecayOP}).
The lifetime is proportional to $(M_{\rm eff.} M_{\rm eff}')^2$.  
Thus, the operators such as (\ref{eq:LSPdecayOP}) are consistent with the 
proton decay experiments as long as the geometric mean of $M_{\rm eff.}$ and 
$M_{\rm eff.}'$ are large enough, say of order of the GUT scale or lager. 

\begin{figure}
\begin{center}
\begin{picture}(400,100)(-200,0)
\DashArrowLine(-160,50)(-100,50){3}
\Text(-165,50)[r]{$\tilde{e^c}$}
\Vertex(-100,50){2}
\Text(-100,45)[rt]{$\vev{h_d}$}
\ArrowLine(-40,80)(-100,50)
\Text(-35,80)[l]{$u^c$}
\ArrowLine(-40,20)(-100,50)
\Text(-35,20)[l]{$q$}
\DashArrowLine(40,50)(100,50){3}
\Text(35,50)[r]{$\tilde{q}$}
\Vertex(100,50){2}
\Text(100,45)[rt]{$\vev{h_d}$}
\ArrowLine(160,80)(100,50)
\Text(165,80)[l]{$q$}
\ArrowLine(160,20)(100,50)
\Text(165,20)[l]{$q$}
\end{picture}
\caption{\label{fig:sqlton-decay}Feymann diagrams of slepton and squark 
decay.}
\end{center}
\end{figure}
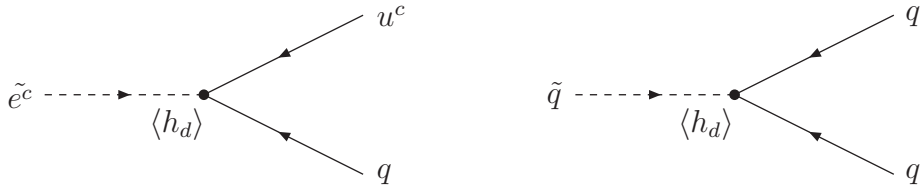

The operators (\ref{eq:LSPdecayOP}) also lead to LSP decay. Either one 
of those operators is enough. Figure~\ref{fig:chi-decay} are some of 
Feynman diagrams contributing to the LSP decay when the LSP is a neutralino.
Sleptons or squarks also decay e.g. through the Feynman diagrams in 
Fig.~\ref{fig:sqlton-decay}), if they are the LSP.
Since the limit from the proton decay only constrains the product of 
$M_{\rm eff.}$ and $M_{\rm eff.}'$, the decay rate of slepton LSP is not 
constrained at all. On the other hand, both operators in (\ref{eq:LSPdecayOP})
contribute to the decay amplitude in the case of neutralino or squark 
LSP, and the decay amplitude may be dominated by amplitudes involving 
only either one of them. 

When the both effective energy scales $M_{\rm eff.}$ and $M_{\rm eff}'$ 
are roughly of the same order, the LSP lifetime is approximately 
\begin{equation}
 \tau \approx \frac{M_{\rm eff.}^2}{(100 \; \GEV)^3} \approx 
1 \; {\rm min.} \; \left(\frac{M_{\rm eff.}}{10^{16} \; \GEV} \right)^2, 
\end{equation}
showing that the LSP decays roughly at the epoch of the big-bang 
nucleosynthesis if $M_{\rm eff.}$ is of order $10^{16}$ GeV.
To be more precise, $M_{\rm eff.}$ and $M_{\rm eff.}'$ have generation 
indices, and those that matter to proton decay and the LSP decay are not 
the same. Since the LSP decay process picks up the largest one, while 
proton decay does not necessarily, $M_{\rm eff}$ that determines the 
LSP lifetime may be even lower than $10^{16}$ GeV, meaning that the LSP 
may decay even before the big-bang nucleosynthesis.
But, the precise lifetime further depends on the mixing among neutralinos, 
for instance, in the case of neutralino LSP, and requires detailed 
calculations. If the LSP decays after the big-bang nucleosynthesis, 
the relic abundance of the LSP (before the decay) is constrained, and 
so is the thermal history of the universe, consequently.

If the lifetime is short enough, say, $c\tau \gamma \lesssim 10 {\rm km}$, 
some of supersymmetry particles produced at the LHC decay inside the 
detectors. The LSP that decays does not contribute to the missing energy.
Jets (and possibly a lepton) come out of a displaced vertex in LSP decay 
events. 
If the lifetime is not that short, the LSP decay outside the detectors 
and we may not notice. But, in the case the LSP is a charged particle, say 
a slepton or a squark, they can be trapped; such experiments have been 
proposed in the context of the NLSP decay to the gravitino in the gravitino 
LSP scenario \cite{gravitino}.
The LSP decay events are completely different from the NLSP decay to 
gravitino, and it would not be difficult to make a distinction between 
them. It would be further interesting if the branching ratios of various 
decay modes of the LSP can be measured, since both $M_{\rm eff.}$ and 
$M_{\rm eff.}'$ can be extracted. We already know that the Yukawa couplings 
of strange quark and muon do not really unify. Thus, the measurement of 
the two energy scales, $M_{\rm eff.}$ and $M_{\rm eff.}'$, may give us 
another clue to understand how the SU(5)$_{\rm GUT}$ unified symmetry 
is broken. Further detailed phenomenological study will be presented 
elsewhere \cite{progress}.

If the LSP lifetime is shorter than the current age of the universe, 
it cannot be a candidate of dark matter. This makes axion an attractive 
candidate of dark matter. The Peccei--Quinn mechanism still remains 
one of the best solutions to the strong CP problem, which predicts an axion 
field. The relic abundance of the axion may be explained by anthropic 
choice of the initial amplitude of the axion filed \cite{Wilczek}. 
It should be reminded that even if the LSP that does not decay within the 
detectors of the LHC, it is still not necessarily stable in the cosmological timescale. 
Indeed, we have seen that it may really be the case in a theoretically 
well-motivated framework.

\section*{Acknowledgements}
We thank Hitoshi Murayama for useful comments.
R.T. thanks the Galileo Galilei Institute for Theoretical Physics for 
the hospitality and the INFN for partial support during the completion 
of this work.
This work was supported in part by a PPARC Advanced Fellowship (R.T.), 
and in part by the Director, Office of Science, 
Office of High Energy and Nuclear Physics, of the US Department of 
Energy under Contract DE-AC02-05CH11231, and in part 
by the Miller Institute for Basic Research in Science (T.W.). 

\appendix 

\section*{Appendix}

In Heterotic string theory with $E_8 \times E_8$ gauge group, a 
rank-5 vector bundle $V_5$ has to be turned on in one of $E_8$ in order to 
obtain an SU(5)$_{\rm GUT}$ unified theory. For the 4+1 and 3+2 models 
of \cite{TW} the $V_5$ is taken at reducible limits such as 
$V_5 = U_4 \oplus L$, where $U_4$ is rank-4 bundle and $L$ a line bundle, 
or $V_5 = U_3 \oplus U_2$, where $U_{3}$ and $U_2$ are rank-3 and -2 
vector bundles, respectively. The structure group of the rank-5 bundle 
is reduced from SU(5) to either SU(4)~$\times$~U(1)$_\chi$ or 
SU(3)~$\times$~SU(2)~$\times$~U(1)$_{\tilde{q}_7}$. The commutant of 
the structure group in $E_8$ is either 
SU(5)$_{\rm GUT}$~$\times$~U(1)$_\chi$ or 
SU(5)$_{\rm GUT}$~$\times$~U(1)$_{\tilde{q}_7}$, the gauge group 
discussed in the main text of this article.

In the 4+1 model, $U_4$-valued (0,1)-form become (scalar part of) 
{\bf 10}-type chiral multiplets, and $\overline{U}_4$-valued (0,1)-form 
become {\bf 10}$^c$-type chiral multiplets. The U(1)$_\chi$ charges 
of {\bf 10}-type chiral multiplets are $-1$, as shown 
in Table~\ref{tab:charge}. The massless modes are $H^1(Z;U_4)$ and 
$H^1(Z;\overline{U}_4)$, where $Z$ is a Calabi--Yau 3-fold for 
compactification. The $\bar{\bf 5}$-type, $\bar{H}(\bar{\bf 5}) = 
H({\bf 5})^c$-type and $\bar{N}$-type chiral multiplets are from 
$U_4 \otimes L$-, $\wedge^2 U_4$- and $U_4 \otimes L^{-1}$-valued 
(0,1)-forms, respectively, and the $\bar{\bf 5}^c$-type, $H({\bf 5})$-type 
and $\bar{N}^c$-type chiral multiplets are from bundles in the Hermitian 
conjugate representation, $\overline{U_4 \otimes L}$, 
$\overline{\wedge^2 U_4}$ and $\overline{U}_4 \otimes L$. 
Note that $\bar{N}^c$-type chiral multiplets were denoted as  
$\overline{\bar{N}}$ in \cite{TW}. 

In the 3+2 model, $U_2$-valued (0,1)-form become {\bf 10}-type chiral 
multiplets, and {\bf 10}$^c$-type multiplets are from the $\overline{U_2}$ 
bundle. The $\bar{\bf 5}$-type, $\bar{H}(\bar{\bf 5})$-type and 
$H({\bf 5})^c$-type chiral multiplets originate from the 
$U_3 \otimes U_2$-, $\wedge^2 U_3$- and $\wedge^2 U_2$ bundle valued 
(0,1)-form, and $\bar{\bf 5}^c$-type, $\bar{H}(\bar{\bf 5})^c$-type 
and $H({\bf 5})$-type chiral multiplets are from their Hermitian conjugate 
bundles. $\bar{N}$-type multiplets are from $\overline{U_3}\otimes U_2$, 
and $\bar{N}^c$-type from $U_3 \otimes \overline{U_2}$.
The massless modes are given by the first cohomology of the 
corresponding vector bundles. The U(1)$_{\tilde{q}_7}$ charges are shown 
in Table~\ref{tab:charge}.

The difference between the $\bar{\bf 5}$-type and 
$\bar{H}(\bar{\bf 5})$-type chiral multiplets is not only due to their different  U(1) charges.
Although we discussed the selection rule that follows only from the 
U(1) symmetry in this article, \cite{TW} discusses the selection rules 
that come from the underlying gauge symmetry SU(4) or SU(3)$\times$SU(2) 
as well. The centre of the structure group, $\Z_4$ and $\Z_3 \times \Z_2$, 
respectively, may not be broken by the gauge connection describing a 
vector bundle $V_5$. In the 4+1 model, the expectation value of 
$\bar{N}^c$-type chiral multiplets leave the diagonal subgroup of 
the matter parity $\Z_{10}/\Z_5 \simeq \Z_2$ and the $\Z_2$ subgroup 
of the centre $\Z_4$ unbroken. But, it turns out that all the chiral 
multiplets are even under this diagonal unbroken $\Z_2$ symmetry, 
and this unbroken symmetry does not have any significance. In the 3+2 
model, the diagonal subgroup of the matter parity and the centre of 
the structure group SU(2) may remain unbroken, but all the chiral 
multiplets are even under this diagonal $\Z_2$ symmetry. Thus, it 
does not lead to a selection rule that is not covered in this article.

Reference \cite{TW} only discussed the wave functions of the zero modes 
in its section 3, because we were interested only in the renormalizable 
part of the effective theory. In this article, we started out with 
field-theory models that contain not only the zero modes but 
also Kaluza--Klein modes and vector-like pair of zero modes 
that do not contribute to the net chirality. In the end, however, what 
we did is to determine the massless modes (zero modes) by diagonalizing 
mass matrices. Thus, the analysis in the former half of this article 
is essentially the same as what we did in section 3 of \cite{TW}. 
The difference is that we used differential equation in internal space and 
geometric intuition in \cite{TW}, while we used truncated D = 4 spectra 
and diagonalization of mass matrices in this article.

%%%%%%%%%%%%%%%%%%%%%%%%%%%%%%%%%%%%%%%%%%%%%%%%%%%%%%%%%%%%%%%%%%%
%%%%%%%%%%%%%%%%%%%%%%%%%%%%%%%%%%%%%%%%%%%%%%%%%%%%%%%%%%%%%%%%%%%

%%%%%%%%%%%%%%%%%%%%%%%%%%%%%%%%%%%%%%%%%%%%%%%%%%%%%%%%%%%%%%
\begin{figure}
\begin{center}
\begin{picture}(400,100)(-200,0)
% \Line(-200,0)(200,0)
% \Line(0,0)(0,100)
%
\ArrowLine(-160,50)(-110,50)
\Photon(-160,50)(-110,50){3}{4}
\Text(-165,55)[rb]{$\tilde{\chi}^0$}
\ArrowLine(-60,90)(-110,50)
\Text(-55,90)[l]{$e^c$}
\DashArrowLine(-110,50)(-80,30){3}
\Text(-90,45)[lb]{$\tilde{e^c}$}
\Vertex(-80,30){2}
\Text(-80,25)[rt]{$\vev{h_d}$}
\ArrowLine(-50,50)(-80,30)
\Text(-45,50)[l]{$u^c$}
\ArrowLine(-50,10)(-80,30)
\Text(-45,10)[l]{$q$}
\ArrowLine(50,50)(100,50)
\Photon(50,50)(100,50){3}{4}
\Text(45,55)[rb]{$\tilde{\chi}^0$}
\ArrowLine(150,90)(100,50)
\Text(155,90)[l]{$q$}
\DashArrowLine(100,50)(130,30){3}
\Text(120,45)[lb]{$\tilde{q}$}
\Vertex(130,30){2}
\Text(130,25)[rt]{$\vev{h_d}$}
\ArrowLine(160,50)(130,30)
\Text(165,50)[l]{$q$}
\ArrowLine(160,10)(130,30)
\Text(165,10)[l]{$q$}
\end{picture}
\caption{\label{fig:chi-decay}Feymann diagrams of neutralino decay.}
\end{center}
\end{figure}
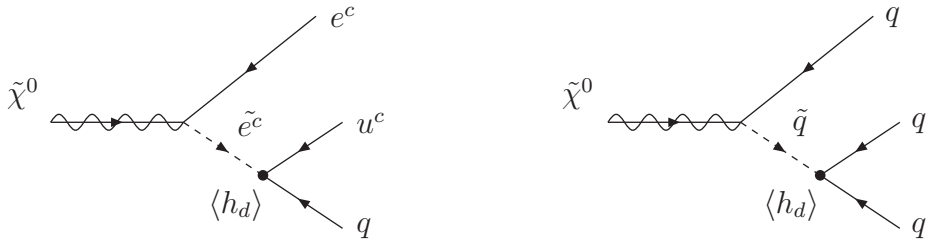
%%%%%%%%%%%%%%%%%%%%%%%%%%%%%%%%%%%%%%%%%%%%%%%%%%%%%%%%%%%%%%

%%%%%%%%%%%%%%%%%%%%%%%%%%%%%%%%%%%%%%%%%%%%%%%%%%%%%%%%%%%%%%
%%%%%%%%%%%%%%%%%%%%%%%%%%%%%%%%%%%%%%%%%%%%%%%%%%%%%%%%%%%%%%%%%%%

\end{document}